\documentclass[aps,twocolumn,amsmath,showkeys,showpacs,superscriptaddress,prb,floatfix]{revtex4-2}
\usepackage{dcolumn}
\usepackage{bm}
\usepackage[T1]{fontenc}
\usepackage{amsmath}
\usepackage{amssymb}
\usepackage{graphicx}
\usepackage{tabularx}
\usepackage{hyperref}
\hypersetup{colorlinks=true, linkcolor=navy, citecolor=navy, urlcolor=navy, pdftitle={Chiral Symmetry and Magnetism in a 3D Kagome lattice: $R$Pt$_2$B ($R$ = La and Nd) prototype crystals}, pdfauthor={A.C.Garcia-Castro}}
\usepackage{xcolor}
\usepackage[mathscr]{eucal}

\usepackage[colorinlistoftodos]{todonotes}

\definecolor{navy}{RGB}{0,0,139}
\definecolor{emerald}{rgb}{0.31, 0.78, 0.47}

\begin{document}
\par\noindent\rule{\textwidth}{0.5pt}
\title{Chiral symmetry and magnetism in a 3D Kagome lattice: $R$Pt$_2$B ($R$ = La and Nd) prototype crystals}

\author{C. E. Ardila-Gutiérrez}
\affiliation{School of Physics, \textsc{ficomaco} Research Group, \href{https://ror.org/00xc1d948}{Universidad Industrial de Santander}, SAN-680002, Bucaramanga, Colombia}

\author{D. Torres-Amaris}
\affiliation{School of Physics, \textsc{ficomaco} Research Group, \href{https://ror.org/00xc1d948}{Universidad Industrial de Santander}, SAN-680002, Bucaramanga, Colombia}
 
\author{Rafael González-Hernández}
\affiliation{Departamento de Física y Geociencias, \href{https://ror.org/031e6xm45}{Universidad del Norte}, ATL-080020, Barranquilla, Colombia.}

\author{Aldo. H. Romero}
\email{aldo.romero@mail.wvu.edu }
\affiliation{Department of Physics and Astronomy, \href{https://ror.org/011vxgd24}{West Virginia University}, WV-26506-6315, Morgantown, United States}

\author{A. C. Garcia-Castro}
\email{acgarcia@uis.edu.co}
\affiliation{School of Physics, \textsc{cimbios} Research Group, \href{https://ror.org/00xc1d948}{Universidad Industrial de Santander}, SAN-680002, Bucaramanga, Colombia}

\thinspace
\date{Received: \today, Revised: XXX, Accepted: XXX, Published: XXX}

\begin{abstract}  
Chirality in crystals arises from the exclusive presence of proper symmetry operations, such as rotations and screw axes, while lacking improper operations like inversion, mirror planes, and roto-inversions. Crystallographic chirality is expected to be coupled with magnetic responses in magnetically active chiral compounds. Therefore, this study investigates the interplay between structural chirality and magnetic ordering in the rare-earth platinum boride family, $R$Pt$_2$B, where $R$ denotes lanthanide elements. Our results show that the $R$ sites structurally form a chiral three-dimensional Kagome lattice, which can lead to magnetic frustration resolved through chiral antiferromagnetic orderings in conjunction with chiral symmetry. Symmetry analysis reveals that these chiral antiferromagnetic states are low-energy states, competing with higher-in-energy (001) ferromagnetic configuration.
We also identified Kramers-type Weyl points in the electronic structure without magnetic response.
In the magnetically active chiral compound NdPt$_2$B, Zeeman splitting lifts degeneracies at the high-symmetry points; however, Weyl points persist due to the breaking of time-reversal ($\mathcal{T}$) and inversion ($\mathcal{P}$) symmetries. We also estimate the anomalous Hall conductivity, a measurable observable of the allowed topological features finding a value of $\sigma_{xy} = 293$ S$\cdot$cm$^{-1}$ comparable with another Kagome magnetic materials like Mn$_3$PtN and FeSn, for example.
This study elucidates the intricate interplay among chirality, magnetism, and topology in rare-earth Kagome materials.
\\
\\
DOI: XX.XXXX/XxxxXxxX.XXX.XXXXXXX
\end{abstract}


\maketitle

\section{Introduction}

The study of chirality across various disciplines, including chemistry, particle physics, and, more recently, condensed matter physics, has unveiled previously hidden couplings and phenomena. These discoveries have become central discussion topics within the scientific community \cite{Fecher2022, bousquet2024structural, Nespolo2021}. 
For example, some works show chiral phonons in chiral symmetry compounds can be proposed as ideal platforms for the design of dark matter detectors \cite{PhysRevResearch.5.043262}, chirality-induced spin selectivity in crystals \cite{doi:10.1021/acs.chemrev.3c00661}, and chirality-induced topological properties in materials thanks to their lack of roto-inversion symmetries \cite{PhysRevB.96.045102,Chang2018,felser_topology_2023,PhysRevB.110.075156}.
The field of chiral materials is experiencing a surge of interest, driven by their potential to revolutionize various industries. Researchers are exploring their applications in designing more efficient electronic circuits, enhancing battery capacities, advancing hydrogen storage solutions, and developing topologically protected magnetic memory devices. These innovations could lead to faster, more energy-efficient technologies, longer-lasting energy storage systems, and robust data storage solutions that leverage the unique properties of chiral structures
\cite{armitage_weyl_2018, hasan_topological_2010, pang_super-stretchable_2016, borisenko_experimental_2014}. 
Among the most commonly studied topological materials are Dirac \cite{young_dirac_2012, young_dirac_2015}, nodal \cite{fang_topological_2016,yan_collective_2016}, and Weyl \cite{burkov_weyl_2018, soluyanov_type-ii_2015, bordoloi2024promises} semimetals. In Dirac semimetals, both time-reversal and inversion symmetries must be present to exhibit fourfold degenerate nodes. The other two types of semimetals may break one or both of these symmetries, leading to two-fold degeneracies.
As such, in these topological materials, structural chirality often plays a crucial role in the emergence of non-trivial behavior \cite{yan_structural_2023,niemann_chiral_2017}. 
Specifically, Weyl semimetals are closely related to chirality in both their crystal structure, where chirality is understood as the property of being non-superposable on the material's mirror image, and electronic dispersion, where it is more related to spin-momentum locking. 
In these systems, the application of parallel external magnetic and electric fields disrupts chirality conservation~\cite{gomez2024structural}, facilitating electronic transport that the chirality of the Weyl nodes would otherwise constrain. Magnetism is pivotal in elucidating the phenomena associated with this process. Therefore, a comprehensive analysis of the complex interactions within these multifunctional systems can provide valuable insights into the fundamental mechanisms governing solid-state physics.
Let us examine how this can be achieved. Topology and chirality have been extensively explored from a crystallographic perspective \cite{you_topological_2023, felser_topology_2023, yan_structural_2023, harris_molecular_1999}. In contrast, magnetism and topology are often studied through quasiparticles such as magnons \cite{shindou_topological_2013, mcclarty2022topological} and skyrmions \cite{rosler_spontaneous_2006}, which emerge from collective phenomena and hold potential for applications in chiral spintronics \cite{NatRev2021}. This macroscopic viewpoint continues to be a critical and active area of research \cite{NatComm2021, Nat0217}. However, there remains a need for approaches that explicitly show how magnetic response, and its induced symmetry considerations, couple to crystallographic symmetry.
As such, in different magnetic orderings, for example in the FM case,  the time-reversal symmetry is lifted, coupling to the crystallographic symmetry operations and leading to changes in its electronic and topological properties.

In this context, chiral 3D Kagome lattices present a valuable opportunity to study these interactions \cite{you_topological_2023}. 
Their atomic structure exhibits a screw axis combined with chiral ordering. Moreover, in magnetically active 3D Kagome systems, spin-momentum coupling naturally leads to triangular frustrated magnetic states \cite{kubler_non-collinear_2014, kuroda_evidence_2017}. In these lattices, only a few spin arrangements are stable, but the symmetry groups for structural and spin arrangements are entangled, allowing several properties to emerge or be canceled out, in terms of the symmetry operations when considering the magnetic orderings and response.
This highlights the necessity of accounting for non-collinear magnetism to achieve accurate modeling. 
This magnetic frustration can lead to nontrivial, non-collinear orderings with antiferromagnetic chiral structures. In these cases, chirality becomes vectorial and arises from mirror symmetry breaking due to magnetic orderings. This behavior is observed in triangular frustrated magnetism, as exemplified by Mn$_3$NiN, which supports the magnetic orderings $\Gamma_{5g}$ and $\Gamma_{4g}$ \cite{Fruchart1978}. Consequently, Kagome lattices emerge as promising candidates to independently host chiral antiferromagnetic ordering or topological features, depending on the dominant interactions. Although materials with chiral symmetry and chiral magnetism in centrosymmetric compounds have been studied separately, the intricate interplay between chiral symmetry and chiral magnetic orderings within a single prototype crystal remains unexplored mainly in detail.

This paper investigates the $R$Pt$_2$B family of crystals, specifically $R$ = La and Nd, as prototypes of chiral crystals with a three-dimensional magnetic Kagome lattice structure. We present a symmetry-driven model based on density functional theory (DFT) calculations to describe these systems' chiral non-collinear magnetic ordering. This model addresses the ongoing debate surrounding experimental measurements, particularly in the magnetically active $R$ = Nd case \cite{PhysRevMaterials.5.034411}. Our findings reveal that the topological features are intrinsically linked to the chiral crystal structure in both cases, $R$ = La and $R$ = Nd. However, we demonstrate that the magnetic response induced by the Nd site significantly enriches the topological landscape, offering unique insights into the interplay between chirality, magnetism, and topology.
Section~\ref{secII} outlines the computational and theoretical methods of this study. Section~\ref{secIII} presents our results, beginning with an analysis of the chiral magnetic ordering allowed by symmetry when $R$ is a magnetically active cation, exemplified by $R$ = Nd. Finally, in Section~\ref{secIV}, we summarize our conclusions and discuss the implications of our findings for future research.

\begin{figure*}[]
        \centering
        \includegraphics[width=18.0cm,keepaspectratio=true]{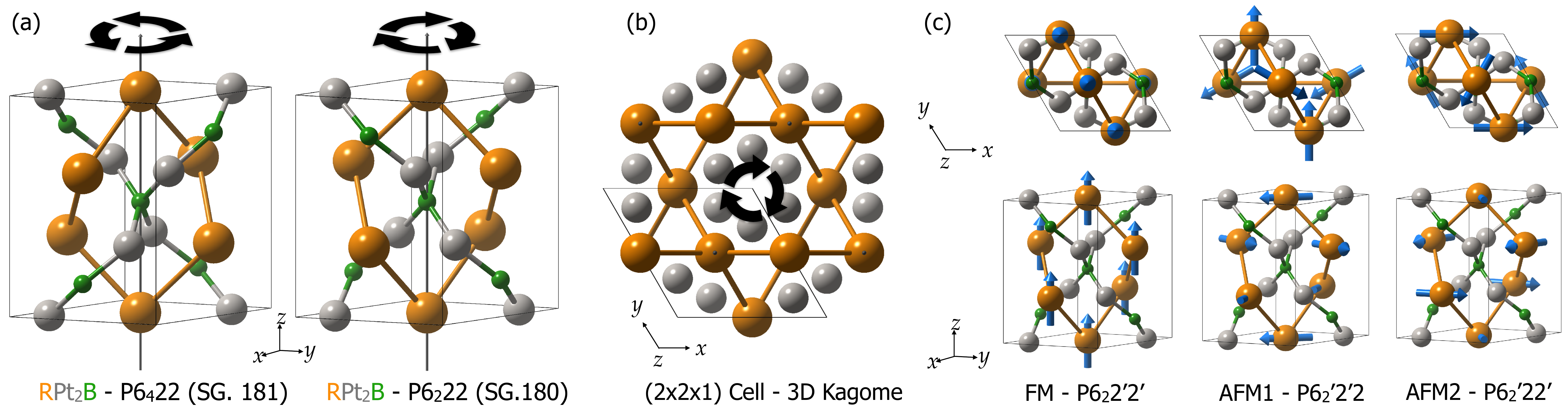}
        \caption{(Color online) (a) $R$Pt$_2$B ($R$ = La and Nd) structure within the SG. 181 - $P6_422$ and SG. 180 - $P6_222$ crystallographic chiral space groups. Here, the screw axis is composed of the Pt-B and $R$-sites as it can be appreciated. (b) $xy$-plane view of the 2$\times$2$\times$1 cell in which, the pseudo-Kagomé 3D chiral lattice can be observed conserving the 3D crystal symmetries. (c) Noncollinear symmetry-allowed magnetic orderings within the primary ferromagnetic FM(001) $P6_22'2'$ (MSG. 180.171) and the antiferromagnetic AFM1 $P6_2$$'2'2$ (MSG. 180.169) and AFM2 $P6_2$$'22'$ (MSG. 180.170) orderings. The latter orderings are allowed when a magnetically active $R$-site is placed, as in the Nd case. The same orderings apply to the SG. 180 - $P6_222$ space group.
        By notation, the $R$, Pt, and B sites are denoted with orange, grey, and green colors, respectively.}
        \label{F1}
\end{figure*}

\section{Computational and theoretical approach:} \label{secII}
We perform first-principles calculations within the density functional theory (DFT) \cite{PhysRev.136.B864,PhysRev.140.A1133} approach using the \textsc{VASP} code (version 5.4.4) \cite{Kresse1996,Kresse1999}. 
The projected augmented wave scheme, PAW \cite{Blochl1994}, represented the valence and core electrons. 
The electronic configurations considered in the pseudopotentials, as valence electrons, are La: (5$s^2$5$p^6$6$s^2$6$d^1$, version 06Sep2000),  Nd: (5$s^2$5$p^6$6$s^2$4$f^4$, version 23Dec2003), B: (2$s^2$2$p^1$, version 06Sep2000), and Pt: (5$p^6$6$s^1$5$d^9$, version 12Dec2005). 
The exchange-correlation was represented within the GGA-PBEsol parameterization of the generalized gradient approximation \cite{Perdew2008}. The Nd:4$f$ orbitals were corrected using the DFT$+U$ approximation within the Liechtenstein formalism \cite{Liechtenstein1995}. 
We used the $U$ = 6.0 eV parameter in the Nd:4$f$ orbitals. This value was selected to properly portray the magnetic and electronic properties of the rare earth elements as reported previously in other compounds \cite{PhysRevB.94.125113,Galler_2022}. 
The periodic solution of the crystal was represented using Bloch states with a Monkhorst-Pack \cite{PhysRevB.13.5188} \emph{k} point mesh of 13$\times$13$\times$11 and 600 eV energy cutoff to give a force convergence of less than 0.001 eV$\cdot$\r{A}$^{-1}$ and an error in energy less than ~10$^{-6}$ eV.  
Spin-orbit coupling (SOC) was included to consider noncollinear magnetic configurations and topological features \cite{Hobbs2000}.  
The anomalous Hall conductivity and associated observables were obtained with the Python package  \textsc{WannierBerri} \cite{tsirkin2021high} using the maximally localized Wannier functions and the tight-binding Hamiltonian generated with the \textsc{Wannier90} package \cite{Pizzi2020}. 
The interpolation was performed with 162 and 174 Wannier functions in LaPt$_2$B and NdPt$_2$B, respectively. The latter with projections on the \textit{d} and \textit{s},\textit{f} orbitals for the La and Nd atoms, \textit{s},\textit{p} for B atoms, and \textit{s},\textit{p},\textit{d} for the Pt sites. For the disentanglement process, we used an energy window $+$4.0 eV above the Fermi level as the maximum, and none for the minimum, and a convergence tolerance of 5.0$\times10^{-8}$ \AA$^2$. 
The illustrations of the atomic structure were created using \textsc{VESTA} software \cite{vesta}.
Finally, the band structure was analyzed with the Python package \textsc{PyProcar} \cite{HERATH2020107080, lang2024expanding}.

\section{Results and Discussion:}\label{secIII}
We compiled a database of experimentally synthesized structures from various sources to investigate the magnetic response in chiral compounds, focusing on those within the 22 enantiomorphic chiral space groups~\cite{bousquet2024structural}. 
Within this search, the chiral crystals within the $R$Pt$_2$B family, with $R$ = La, Y, Ce, Gd, and Nd \cite{PhysRevMaterials.6.104412, PhysRevB.107.214420,SOLOGUB200040,PhysRevMaterials.5.034411}, stand out thanks to their chiral Pt$_2$B chains with chiral rotation along the $z$-axis and embedded into the hexagonal 3D Kagomé lattice composed of the $R$ sites; see Fig. \ref{F1}(a).
The enantiomeric pair is the space groups $P6_222$ (SG. 180) and $P6_422$ (SG. 181).
In the $P6_222$ space group, the $R$-, Pt-, and B-sites occupy the $3d$, $6j$, and $3c$, respectively, with positions (0.5, 0.0, 0.5), (0.151, 0.302, 0.5), and (0.5, 0.0, 0.0).
In the enantiomorph space group $P6_422$, the $R$-, Pt-, and B-sites occupy the $3c$, $6i$, and $3d$, respectively, and are located at the positions (0.5, 0.0, 0.0), (0.151, 0.302, 0.5), and (0.5, 0.0, 0.5) in the hexagonal unit cell.
We analyzed and classified the symmetry-allowed magnetic orderings for a magnetically active $R$-site. This required examining the permissible symmetry operations within the  $P6_222$ and $P6_422$  space groups.
Focusing on the 3$d$ (for the SG. 180) and 3$c$ (for the SG. 181) Wyckoff positions as designated magnetic sites, we applied the corresponding symmetry operations and transformations to these sites, following the approach outlined in Ref. \cite{Perez-Mato_2016} and described in the $k$-Subgroupsmag Tool from the Bilbao Crystallographic Server. 
Within the \textbf{q} = (0,0,0) propagation wave vector, we identified three symmetry-independent magnetic orderings, as illustrated in Fig. \ref{F1}(c), that preserve the parent space group. These consist of a $z$-axis oriented ferromagnetic (FM001) ordering with a magnetic space group (MSG. 180.171) $P6_22'2'$ and two $xy$-plane antiferromagnetic (AFM) orderings with $P6_2$$'2'2$ (MSG. 180.169, labeled as AFM1) and $P6_2$$'22'$ (MSG. 180.170, labeled as AFM2), respectively. 
The latter was found similarly for the SG. 181 with the magnetic space groups as $P6_42'2'$ (MSG. 181.177), $P6_4$$'2'2$ (MSG. 181.175), and $P6_4$$'22'$ (MSG. 181.176) for the FM001 and two chiral in-plane AFM1 and AFM2 orderings.
In particular, the $xy$ plane projections of these magnetic orderings in the 3D Kagome lattice closely resemble those observed in the 2D Kagom\'e structure, such as those found in the [111] family of planes in magnetically active antiperovskites \cite{doi:10.1143/JPSJ.44.781,PhysRevB.106.195113,PhysRevMaterials.6.125003} and in the 2D Kagom\'e compound KMn$_3$Sb$_5$ \cite{PhysRevB.108.245143}.
Moreover, the magnetic ordering handiness is characterized by $k=\frac{2}{3\sqrt{3}}\sum_{ij}[\vec{S_i}\times\vec{S_j}]$, where $i,j$ represents all magnetic moments within the defined unit cell \cite{grohol_spin_2005,PhysRevB.78.144404}. This relationship determines the chirality of noncollinear antiferromagnetic orderings, and for instance, the MSGs $P6_2$$'22'$ and MSG. $P6_2$$'2'2$ in SG.180 exhibits a chiral $+1$ magnetic type.

After these considerations, we comprehensively analyzed the atomic and electronic structures for both magnetic and non-magnetic prototypes within SG. 180. For the non-magnetic LaPt$_2$B system, full atomic and electronic relaxations yielded lattice parameters of $a$ = 5.515 \r{A} and $c$ = 7.835 \r{A}, which closely align with the experimentally measured values of $a$ = 5.522 \r{A} and $c$ = 7.898 \r{A} \cite{SOLOGUB200040}.
In the magnetically active NdPt$_2$B chiral crystal, assuming collinear ferromagnetic ordering as the initial reference, our calculations yielded lattice parameters of $a$ = 5.424 \r{A} and $c$ = 7.880 \r{A}. These results closely match the experimentally observed values of $a$ = 5.4426 \r{A} and $c$ = 7.9001 \r{A} \cite{SOLOGUB200040, PhysRevMaterials.5.034411}.
The same analysis was conducted for the enantiomorphic structure of SG. 181. As expected, no significant differences were observed in the lattice parameters, atomic positions, or magnetic moments for NdPt$_2$B.
In the case of $R$ = Nd, the magnetic response has its origin in the 4$f$:Nd sites, which contain three localized electronic sites at this orbital in the Nd$^{3+}$ state, achieving a $m$ = 2.967 $\mu_B$ magnetic moment per Nd atom, after complete relaxation under non-collinear magnetic ordering. 
Due to the strong electronic correlations in the NdPt$_2$B compound, structural optimization faces significant convergence challenges when the Coulomb exchange correction (Hubbard $U$ term) is omitted. Without this correction, the 4$f$ orbitals contribute heavily to the electronic states near the Fermi level, underscoring the strongly correlated nature of these crystals.

\begin{figure*}[]
        \centering
        \includegraphics[width=18.0cm,keepaspectratio=true]{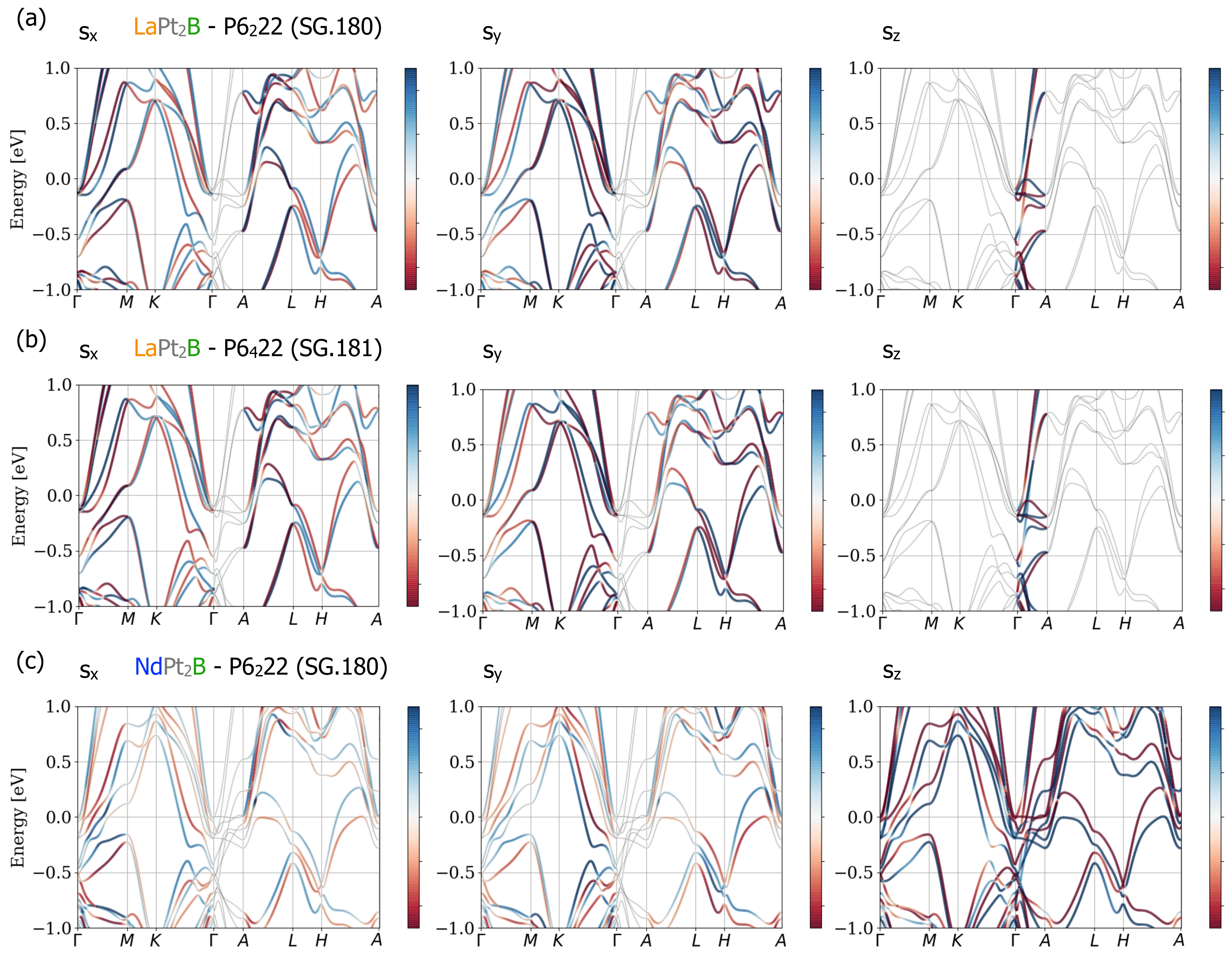}
        \caption{(Color online) Spin-polarized electronic band structure of the LaPt$_2$B in its SG.180, in (a), and at the enantiomorph SG.181, in (b).  It also presented the band structure for the NdPt$_2$B SG.180, in (c). For all the figures, spin polarization is defined in blue and red for spin-up and spin-down projections, respectively. In the case of LaPt$_2$B, we observe that although there is no presence of net magnetic moment in the structure, there is a breaking of the spin-degeneracy induced by the breaking of inversion-symmetry as a result of the chiral space group-related symmetry operations. The latter spin-polarization is reversed due to the reversal of the screw-axis symmetry operation in the SG. 181. In the NdPt$_2$B case, in addition to the inversion symmetry breaking, as in the previous case, we count with the breaking of the time-reversal symmetry induced by the ferromagnetic ordering along the (001) axis as a consequence of the magnetically-active Nd-site occupying the $3d$ Wyckoff positions.}
        \label{F2}
\end{figure*}

Upon closer examination, both right-handed $R$Pt$_2$B ($R$ = La and Nd) SG. 180 ternary compounds exhibit nearly identical crystalline structures and helicoidal arrangements of the atomic sites. Alternatively, the left-handed NdPt$_2$B SG. 181, the mirror image of the NdPt$_2$B SG. 180, shows an inverted helix compared to its counterpart; see Figs. \ref{F1}(a) and \ref{F1}(b).
In agreement with the experimental reports \cite{SOLOGUB200040, PhysRevMaterials.5.034411}, NdPt$_2$B and LaPt$_2$B, have fairly similar lattice parameters. However, the data presented so far indicate that there are no major differences between LaPt$_2$B and NdPt$_2$B other than the substitution of La by Nd. 
The magnetic nature of Nd significantly influences the system’s electronic properties, shaping them in relation to its magnetic structure, as further discussed in this paper.
Figure \ref{F1}(c) illustrates the configurations of chiral magnetic structures allowed by symmetry. Notably, when viewing the $z$-axis, the in-plane magnetic configurations in Fig. \ref{F1}(c) resemble the $\Gamma_{4g}$ and $\Gamma_{5g}$ chiral antiferromagnetic orderings observed in antiperovskites and Kagome lattices, as previously discussed.
A closer examination from a standard perspective reveals that these magnetic orderings are distributed across multiple planes at different $z$-positions. Notably, the structure shown in Fig. \ref{F1}(a), which resembles a blender blade, preserves only the screw axis symmetry along the $z$-direction, eliminating all other symmetry operations.
Each magnetic ordering can be interpreted as a combination of in-plane collinear ferromagnetic layers, with each successive layer in the $xy$-plane rotated by 120 degrees along the $z$-direction.
As these are primary noncollinear magnetic orderings, those can also be combined in symmetry-allowed states, for the same \textbf{q} = (0,0,0) propagation vector, where the chiral in-plane AFM is added to the out-of-plane FM ordering. The latter combinations resulted in the $P3_12'1$ (MSG. 152.35) and $P3_112'$ (MSG. 151.31) for the AFM1+FM001 and AFM2+FM001 cases, respectively, when considering the SG. 181 space group. If the parent space group considered is the SG. 180, the $P3_22'1$ (MSG. 154.43) and $P3_212'$ (MSG. 153.39) for the AFM1+FM001 and AFM2+FM001 cases are obtained. 
Notably,  $P3_112$ (SG 151) and  $P3_212$ (SG 153) belong to one of the 11 enantiomorphic pairs of crystallographic space groups, alongside  $P3_121$ (SG 152) and  $P3_221$  (SG 154).
As these magnetic orderings and magnetic space groups are the result of the combinations of the primary orderings, we focus our efforts on the FM, AFM1, and AFM2 orderings.
After full atomic, electronic, and non-collinear magnetic relaxation, we observed that in terms of the total energy per formula unit, the lowest energy magnetic state belongs to the chiral AFM2 ($P6_2$$'22'$)  ordering followed by the AFM1 ($P6_2$$'2'2$) and the FM001 ($P6_22'2'$) magnetic ordering.  
Here, the energy differences are $\Delta E$ = $-$292.6 meV$\cdot$f.u.$^{-1}$ and $\Delta E$ = $-$290.1 meV$\cdot$f.u.$^{-1}$ for the $\Delta E$ = $E_{AFM2}-E_{FM001}$ and $\Delta E$ = $E_{AFM1} - E_{FM}$, respectively.
Our analysis offers valuable insight into the experimentally observed H-T diagrams, where an S-shaped behavior has been identified. This behavior has been tentatively linked to a Heisenberg model describing triangular antiferromagnetic (AFM) and Kagom\'e lattices \cite{PhysRevMaterials.5.034411,Gvozdikova_2011,PhysRevB.84.214418}.
Furthermore, as noted in \citeauthor{PhysRevMaterials.5.034411}, describing the crystal structure and spin configurations of NdPt$_2$B as a simple two-dimensional triangular AFM is challenging. Nevertheless, given that nonmagnetic boron layers separate the magnetic Nd layers, the two-dimensional magnetic interactions might play a dominant role in determining the main characteristics of the complex magnetic response of NdPt$_2$B \cite{PhysRevMaterials.5.034411} and therefore, our results reveal pseudo-in-plane chiral antiferromagnetic orderings as the lower-energy states.

To elucidate the magnetic contributions of the Nd sites in  $R$Pt$_2$B compounds, we first analyze the electronic properties of the non-magnetic LaPt$_2$B prototype. The spin-projected band structure, presented in Fig. \ref{F2}(a), reveals that LaPt$_2$B exhibits metallic behavior, with multiple bands crossing the Fermi level along key high-symmetry directions in the hexagonal Brillouin zone, particularly along the  $L$–$H$–$K$–$M$ path. Establishing these fundamental electronic properties of LaPt$_2$B is crucial for isolating and understanding the magnetic effects introduced by Nd in the  $R$Pt$_2$B series.
In Fig. \ref{F2}(a), the band structure of LaPt$_2$B in space group 180 includes the spin-orbit coupling (SOC) interaction. Notably, the strong SOC, combined with the chiral symmetry of the crystal, lifts the spin degeneracy of the electronic states along certain high-symmetry $k$-paths, depending on the projected spin ($s_x$, $s_y$, and $s_z$) components.

The observed degeneracy breaking is attributed to the fact that LaPt$_2$B and NdPt$_2$B crystals belong to the $622$  point group, which lacks inversion symmetry. However, the additional lifting of degeneracy induced by spin-orbit coupling (SOC) suggests that the emergence of Weyl nodes and associated phenomena, such as anomalous Hall conductivity (AHC), is inherently linked to this interaction \cite{annurev:/content/journals/10.1146/annurev-conmatphys-031016-025458}.
The emergence of Weyl nodes in these compounds necessitates the concurrent breaking of either inversion ($\mathcal{P}$) or time-reversal ($\mathcal{T}$) symmetries, in addition to the presence of non-symmorphic symmetries, and the inclusion of spin-orbit coupling (SOC) \cite{annurev:/content/journals/10.1146/annurev-conmatphys-031016-025458}. 
As expected for enantiomers of chiral compounds, the non-spin-polarized electronic band structures of LaPt$_2$B in SG 180 (Fig. \ref{F2}(a)) and SG 181 (Fig. \ref{F2}(b)) are indistinguishable. However, their distinction emerges in the spin-projected band structures, where a fully inverted spin texture is observed, as indicated by the reversed spin projections. This inversion arises from the opposite screw-axis symmetries in the  $P6_222$ and $P6_422$  space groups.
Interestingly, in both enantiomorphs, the $s_z$ spin component appears exclusively along the $\Gamma$--$A$ direction, which coincides with the $6_2$ and $6_4$ screw symmetry axes of the SG. 180 and SG. 181 space groups, respectively. This arises because the $z$-spin component commutes with the six-fold screw symmetry along the $\Gamma$--$A$ path.
Furthermore, the Weyl nodes are predominantly concentrated along the $\Gamma$--$A$ path, where the screw-axis symmetry induces a phase shift in the eigenvectors of the six-fold rotation from $\Gamma$ to $A$, resulting in symmetry-protected Weyl points, as previously reported by other authors \cite{Chang2018,PhysRevMaterials.4.124203}. This is clearly illustrated in Fig. \ref{F2A} (Appendix \ref{secB}), where multiple Weyl points along the $\Gamma$--$A$ direction are observed within an energy window ranging from approximately $-$0.5 eV to $+$0.5 eV relative to the Fermi level.

The electronic structure obtained in the FM001 magnetically active NdPt$_2$B, as shown in Fig. \ref{F2}(c), on the other hand, also displayed a metallic response but with a time-reversal symmetry breaking induced by the magnetism lifting the spin degeneracy at the high-symmetry points, affecting the Kramers-Weyl nodal points \cite{massive-fermions-2018,NPJ-2023,Chang2018}. 
As expected for the enantiomorphic pair SG 181 in the NdPt$_2$B, in Fig. \ref{F1A}(a) in Appendix \ref{secA}, the full reversal of the spin-polarization is achieved in the electronic structure as a consequence of the reversal of the screw axis when compared to the SG. 180 space group.
In the case of the AFM1 and AFM2 magnetic orderings, presented in Figs. \ref{F1A}(b) and \ref{F1A}(c) in Appendix \ref{secA}, the electronic band structure showed mixed features as, for example, the $s_z$ spin splitting is preserved along the $\Gamma$--A screw axis meanwhile spin degeneracy breaking is noted in the $k_z$=0 plane. This is due to the in-plane ferromagnetic ordering, as exposed previously, and the frustrated AFM coupling between the $xy$ planes stacked along the $z$-axis.

\begin{figure}[t]
        \centering
        \includegraphics[width=8.7cm,keepaspectratio=true]{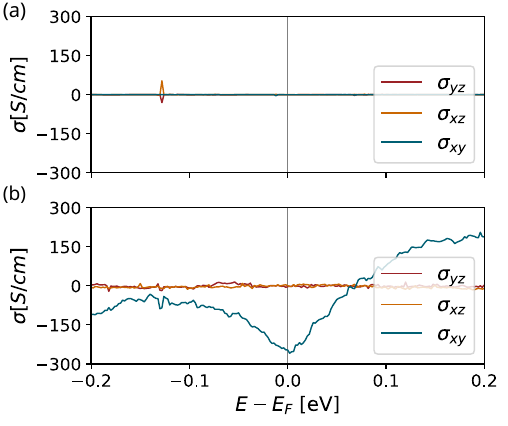}
        \caption{(Color online) Anomalous Hall conductivity, $\sigma_{ij}$, computed in the LaPt$_2$B case, in (a), as well as in the NdPt$_2$B ferromagnetic (001) ground state, in (b). All the $\sigma_{ij}$ components vanished in the non-magnetic case in agreement with the   $\mathcal{T}$-symmetry preserved. In contrast, non-zero $\sigma_{xy}$ is obtained for NdPt$_2$B as expected for the $\mathcal{T}$ symmetry breaking induced by the ferromagnetic ordering with easy-axis along the $z$ direction. The latter agrees with Eq. \ref{ahc:tensor} obtained from symmetry considerations.}
        \label{F4}
\end{figure}

We then investigated the potential topological properties, focusing on anomalous Hall conductivity (AHC). A symmetry-based analysis revealed that, despite their magnetically active nature, the AHC is symmetry-forbidden in the AFM1 and AFM2 magnetic states. This contrasts sharply with antiperovskite Kagome lattices, where the presence of mirror symmetry ($\mathcal{M}$) combined with time-reversal symmetry ($\mathcal{T}$) enables a nonzero Berry curvature, leading to a finite AHC in the chiral antiferromagnetic $\Gamma_{4g}$ phase \cite{PhysRevMaterials.3.044409}.
In the FM001 case, as expected, $\mathcal{T}$ symmetry is broken and therefore, the symmetry-imposed conductivity tensor in the case of NdPt$_2$B (MSG. 180.171) with magnetic point group $6_2$$'2'$ is given as \cite{Gallego:lk5043,PhysRevB.92.155138,PhysRevB.23.2824}:

\begin{align}\label{ahc:tensor}
\sigma_{P6_22'2'}=
\begin{pmatrix}
0 & \sigma_{xy} & 0\\
-\sigma_{xy} & 0 & 0 \\
0 & 0 & 0
\end{pmatrix}
\end{align}

We estimated the anomalous Hall conductivity allowed by symmetry in the NdPt$_2$B FM001, as shown in Fig. \ref{F4} following the procedure provided by the Kubo formula \footnote{The $\sigma_{xy}$  anomalous Hall conductivity component was computed by the formula:
 \begin{align}\label{eq:ahc}
   \sigma_{xy}=-\frac{e^2}{\hbar}\int_{BZ}\frac{d^3\bf{k}}{(2\pi)^3}\Omega_{xy}(\bf{k}),
\end{align}
where $\Omega_{xy}(\bf{k})$=$\sum_n^{occ} f_n(\bf{k})$$\Omega_{n,xy}(\bf{k})$ is the summation of all the occupied $n$ number of bands and $f_n\bf(k)$ is the Fermi distribution.}\cite{PhysRevLett.49.405}. 
The $R$ = La case was also computed and included for comparison and confirmation, see Fig. \ref{F4} (a). This aims to prove its disappearance in the absence of $\mathcal{T}$ symmetry breaking induced by the magnetic response despite the presence of the Weyl nodes, as presented before. 
The latter is due to the cancellation of the Berry curvature, in the entire BZ, induced by the opposite chiralities, in agreement with $\Omega_{ij}(\textbf{k})$ = $-$$\Omega_{ij}(-\textbf{k})$ \cite{NIELSEN198120,NIELSEN1981219}. 
In the NdPt$_2$B compound, the $\sigma_{xy}$ value was found as 293 S$\cdot$cm$^{-1}$.
Despite the zero values of the AHC in the chiral AFM orderings, we believe that the AFM coupled to crystalline chirality could reveal promising features of these magnetic orderings to influence physical properties. Thus, a more in-depth investigation is needed and will be explored in future studies.

\begin{figure}[]
        \centering
        \includegraphics[width=8.8cm,keepaspectratio=true]{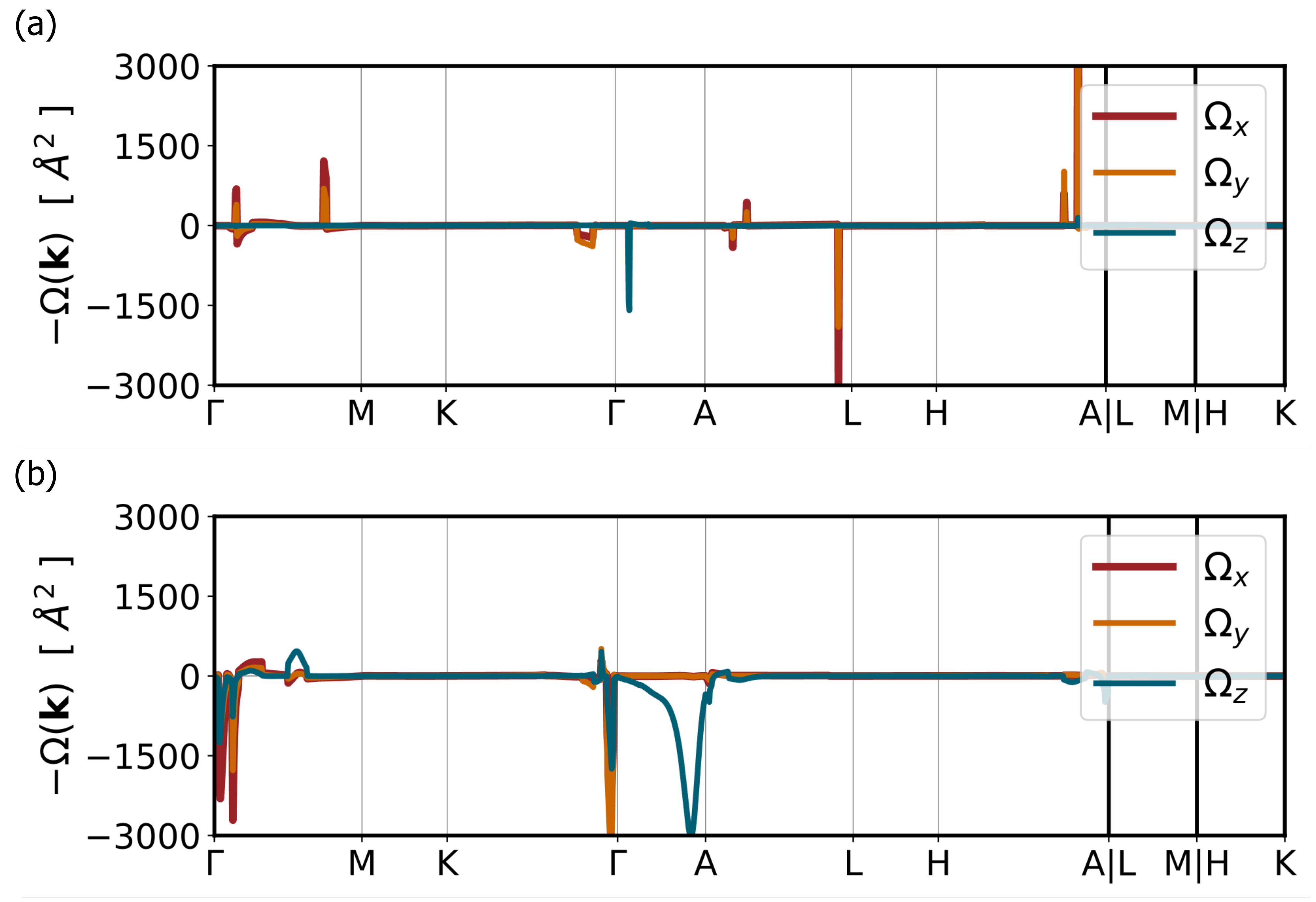}
        \caption{(Color online) Berry-curvature obtained for both, LaPt$_2$B in (a) and NdPt$_2$B in (b) integrated along the high-symmetry BZ path associated with the SG. 181 space group.  As it can be appreciated in the LaPt$_2$B case, only divergent values of the BC are appreciated associated with Weyl nodal crossings. This BC picks with opposite BC values induces a total cancellation when integrating over the entire BZ. (b) In the NdPt$_2$B it can be noted the non-divergent $\Omega_z$ component with a pronounced value along the $\Gamma$--$A$ path and explained in terms of the strong bands repulsion and avoided crossings close to the Fermi energy,  observed in this direction parallel to the screw-axis.}
        \label{F3}
\end{figure}

To better understand the topological properties of the non-magnetic and magnetic chiral crystals LaPt$_2$B and NdPt$_2$B, respectively, in Fig. \ref{F3} we present the estimated Berry curvature, $\Omega_{ij}$. The Berry curvature was obtained for the high-symmetry path along the BZ.
Here, it can be observed that in the case of LaPt$_2$B in Fig. \ref{F3} (a), the Berry curvature is dominated by divergent peaks characteristic of band crossings; see also Fig. \ref{F3A}. We notice, as expected due to the inversion-symmetry breaking in the presence of time-reversal symmetry, that $\Omega_{ij}(\bf{k})$ = $-\Omega_{ij}(\bf{-k})$ and therefore the complete cancelation in the BZ integral is obtained when estimating the AHC. 
However, in the FM001 NdPt$_2$B case in Fig. \ref{F3}(b) in Appendix \ref{secB}, both $\mathcal{P}$ and $\mathcal{T}$ are broken, and thus a non-vanishing value of the Berry curvature is expected. Moreover, as noted in the $\Gamma$ -- $A$ path, there is a tangible presence of the $\Omega_z$ component with a non-divergent character characteristic of degeneracy lifting and avoided bands' crossings close to the Fermi level \cite{kubler_non-collinear_2014,PhysRevB.76.195109,PhysRevLett.92.037204}, in agreement with the $E_{SOC}+E_{Zeeman}$ split bands in the Fermi energy level, see Fig. \ref{F4A} in Appendix \ref{secB}.

Finally, it is crucial to recognize that the interplay between the crystalline and magnetic structures in this chiral compound plays a fundamental role in understanding its topological features. For example, as observed in the chiral antiferromagnetic stable states, in Fig. \ref{F2A} in the Appendix \ref{secA}, the Weyl nodal crossings are still allowed for several bands, as in the M-point in an energy close to -0.2 eV but Zeeman split in the same point at an energy close to 0.7 eV. Therefore, further studies in the AFM chiral states are worth exploring.

\section{Conclusions and general remarks:}\label{secIV}
We have theoretically investigated the chiral compounds NdPt$_2$B (magnetic) and LaPt$_2$B (non-magnetic) as representative prototypes of the $R$Pt$_2$B family of materials. This study was based on first-principles calculations within the density functional theory (DFT) framework, combined with Wannier functions-based analysis.
Here, we explored the chiral structure as a 3D Kagome lattice that allows for the chiral non-collinear antiferromagnetic orderings and 001 ferromagnetic states for a magnetically active $R$-site, $R$ = Nd.
Based on symmetry considerations, we have proposed two chiral antiferromagnetic structures compatible with the crystal's chiral symmetry, which, in this case, resembles the ordering observed in 2D Kagome lattices.
The latter states shed light on the magnetic structures associated with the recent experimental findings on the NdPt$_2$B crystal.
In the non-magnetic case, LaPt$_2$B, the chiral symmetry introduced by the chiral space group induces the presence of Kramers-Weyl points as topological features.
In the magnetically active case, $R$ = Nd, the ferromagnetic state allows the appearance of a tangible anomalous Hall conductivity of a value here estimated to be $\sigma_{xy}$ = 293 S$\cdot$cm$^{-1}$.
Thus, we hope that our results will help the search for novel crystals where chirality, magnetism, and topological features entangle to give rise to future next-generation technologies.
 
\section*{Acknowledgements}
Calculations presented in this article were carried out using the GridUIS-2 experimental testbed, developed in the High Performance and Scientific Computing Center of the Universidad Industrial de Santander (SC3-UIS), development action with the support of the UIS Vicerrectora de Investigación y Extension (VIE-UIS) and several UIS research groups, as well as other funding resources.
We acknowledge the Pittsburgh Supercomputer Center (Bridges2) and the San Diego Supercomputer Center (Expanse) for computational resources provided through allocation DMR140031 under the Advanced Cyberinfrastructure Coordination Ecosystem: Services $\&$ Support (ACCESS) program, supported by NSF grants $\#$ 2138259, $\#$ 2138286, $\#$ 2138307, $\#$ 2137603, and $\#$ 2138296.
Additionally, we appreciate the computational support from the WVU Research Computing Dolly Sods HPC cluster, partially funded by NSF OAC-2117575.
We also acknowledge funding from the West Virginia Higher Education Policy Commission through the Research Challenge Grant Program 2022 (RCG 23-007 Award) and NASA EPSCoR Award 80NSSC22M0173.
A.C.G.C. acknowledges the computational support Laboratorio de Supercomputo del Sureste (LNS), Benemérita Universidad Autónoma de Puebla, BUAP, extended to us for performing heavy theoretical calculations.
A.C.G.C. acknowledge the grant No. 4211 entitled “Búsqueda y estudio de nuevos compuestos antiperovskitas laminares con respuesta termoeléctrica mejorada para su uso en nuevas energías limpias” supported by Vicerrectoría de Investigaciones y Extensión, VIE--UIS.

\appendix

\section{Spin-projected band structure for the noncollinear AFM orderings in the NdPt$_2$B:}\label{secA}

\begin{figure*}[]
        \centering
        \includegraphics[width=18.0cm,keepaspectratio=true]{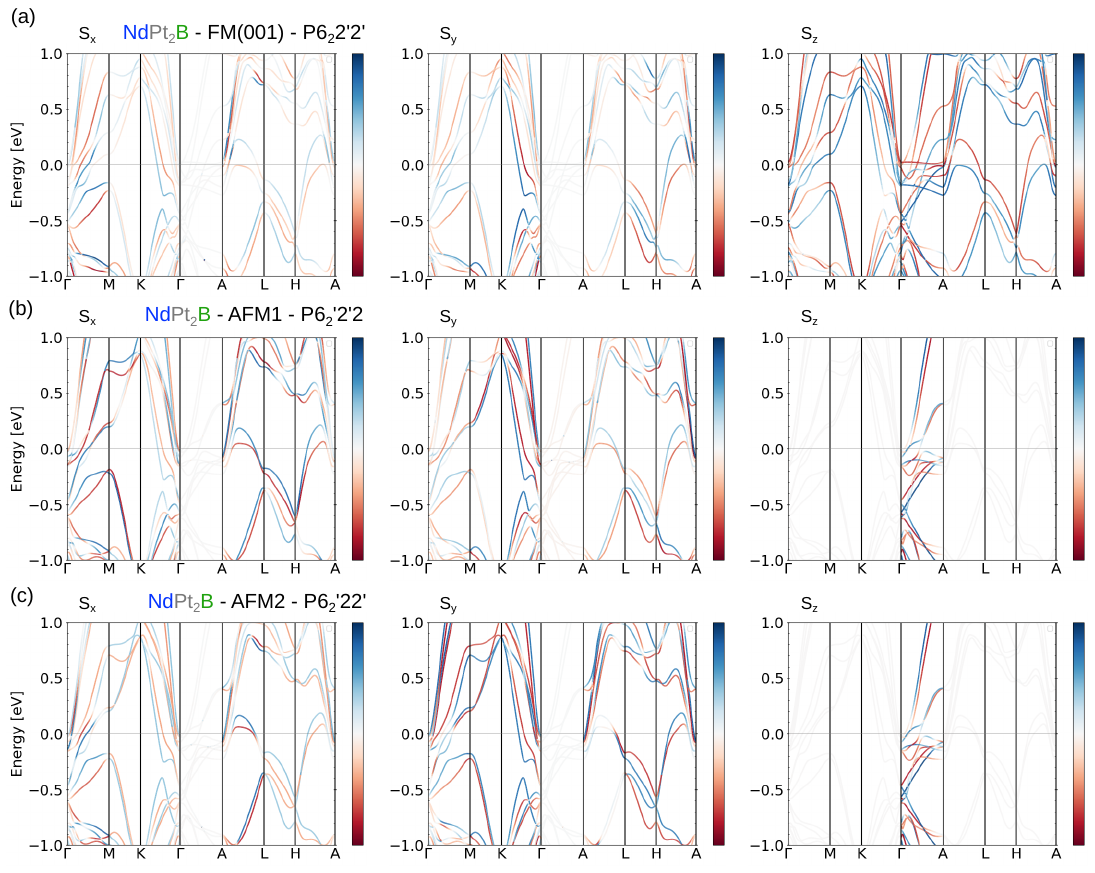}
        \caption{(Color online) (a) Spin-polarized electronic bandstructure computed for the SG.181 space group in the FM(001) down magnetic moment orientation in the NdPt$_2$B magnetically active chiral compound. In (b) and (c) we present the spin-polarized band structure considering the AFM1 and AFM2 chiral orderings related to the MSG. $P6_2$$'2'2$ and MSG. $P6_2$$'22'$, respectively. Here, the $s_x$, $s_y$, and $s_z$ spin components are shown in blue and red colors for the up and down orientations, respectively.}
        \label{F1A}
\end{figure*}

To portray the influence of the magnetic ordering in the electronic structure of the magnetically active chiral NdPt$_2$B, see Fig. \ref{F1A}, we computed the spin-polarized electronic band structure considering the SG.181 in the FM(001) down magnetic moment orientation, in Fig. \ref{F1A}(a). Moreover, we also computed and presented the spin-polarized band structure considering the AFM1, in Fig. \ref{F1A}(b), and AFM2, in Fig. \ref{F1A}(c), chiral antiferromagnetic orderings related to the MSG. $P6'_222'$ and MSG. $P6'_22'2$, respectively. Here, the $s_x$, $s_y$, and $s_z$ spin components are shown in blue and red colors for the up and down orientations, respectively.

\section{Observed Weyl nodes in the LaPt$_2$B and NdPt$_2$B:}\label{secB}

\begin{figure}[]
        \centering
        \includegraphics[width=8.6cm,keepaspectratio=true]{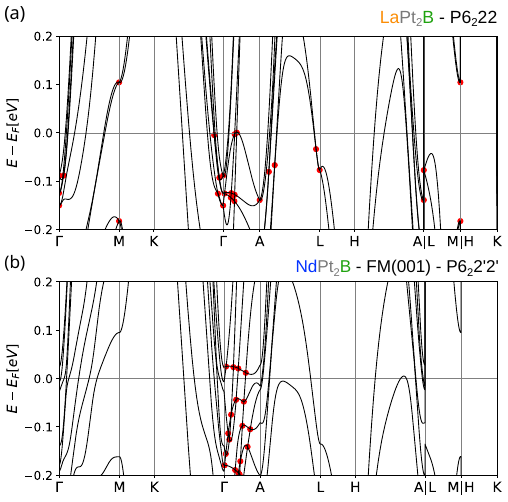}
        \caption{(Color online) Observed Weyl nodes in the spin-polarized electronic bandstructure computed for the SG.181 space group in the LaPt$_2$B, in (a), as well as in the FM(001) magnetic moment orientation in the NdPt$_2$B magnetically active chiral compound, in (b).}
        \label{F2A}
\end{figure}

By employing the symmetry analysis, as implemented in the \textsc{IrReps} toolkit \cite{IRAOLA2022108226}, we extracted the Weyl nodes, as well as their chirality in the LaPt$_2$B and NdPt$_2$B, see Fig. \ref{F2A}. In the nonmagnetic case, the Kramers-Weyl nodal crossings are observed as a consequence of the chiral symmetry. In the ferromagnetic NdPt$_2$B, the nodal points are gapped in the high-symmetry points due to the Zeeman field in the compound \cite{NPJ-2023,massive-fermions-2018}. Nevertheless, other Weyl points are observed in between the $\Gamma$ -- A path due to the $\mathcal{T}$ and $\mathcal{P}$ breaking in the NdPt$_2$B case.

\begin{figure}[]
        \centering
        \includegraphics[width=8.6cm,keepaspectratio=true]{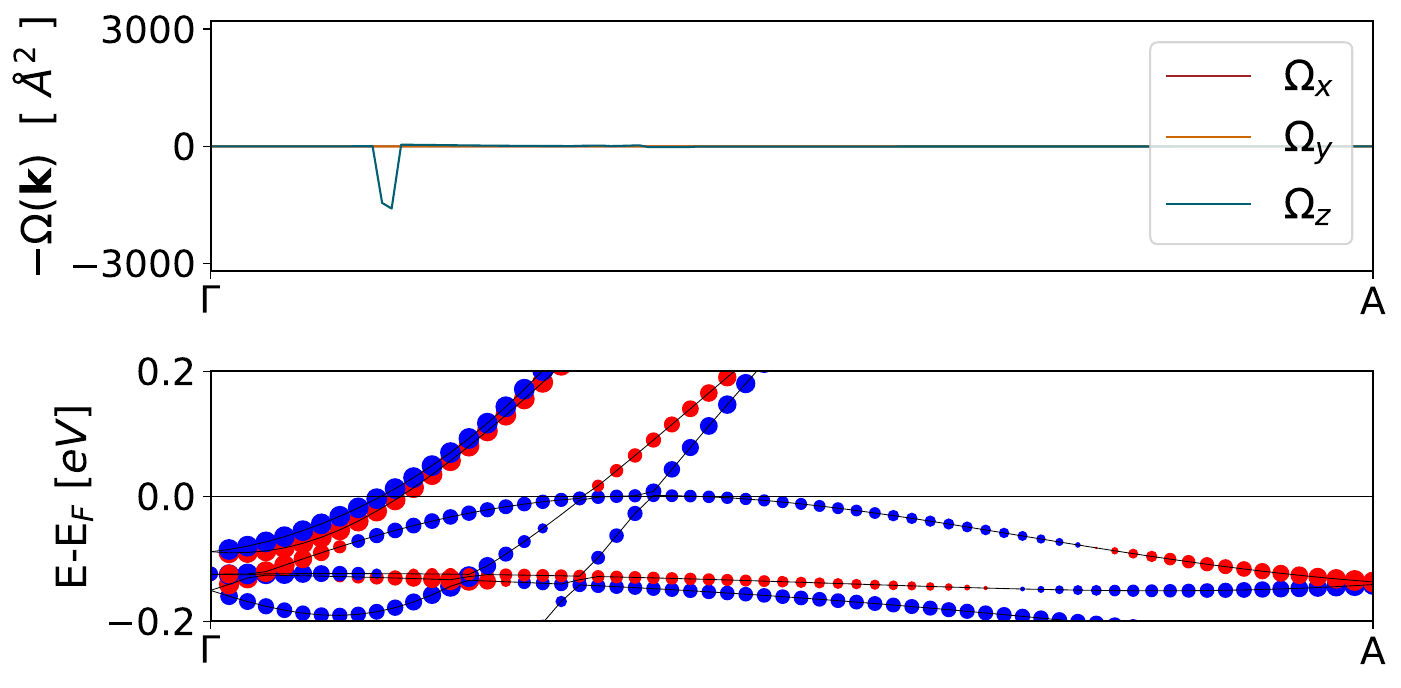}
        \caption{(Color online) (a) Integrated Berry curvature along the high-symmetry $\Gamma$ -- A screw $z$-axis. (b) $s_z$ component spin-projected electronic band structure along the high-symmetry $\Gamma$ -- A path. Both analyses were obtained for the LaPt$_2$B crystal.}
        \label{F3A}
\end{figure}

\begin{figure}[]
        \centering
        \includegraphics[width=8.6cm,keepaspectratio=true]{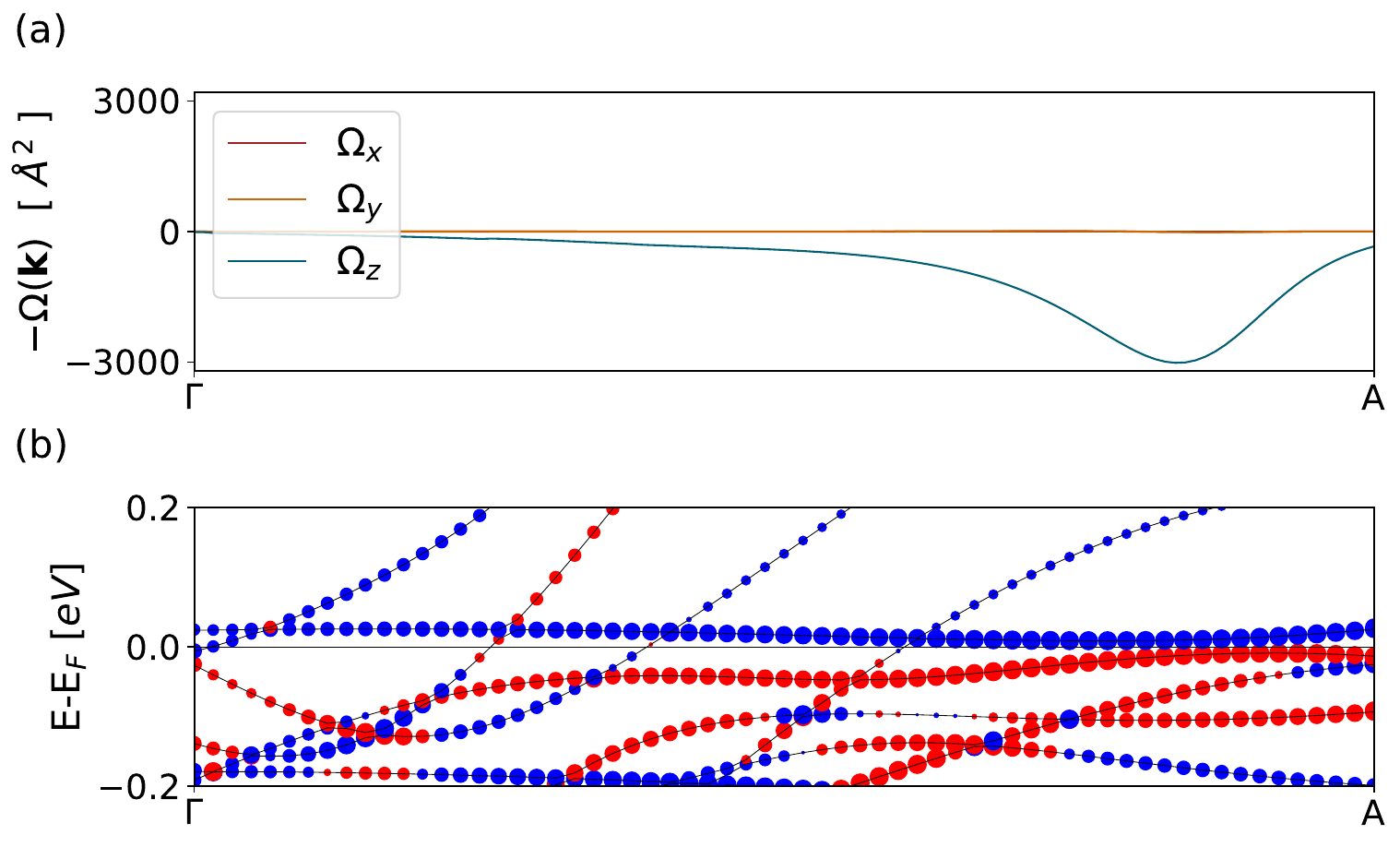}
        \caption{(Color online) (a) Integrated Berry curvature along the high-symmetry $\Gamma$ -- A screw $z$-axis. (b) $s_z$ component spin-projected electronic band structure along the high-symmetry $\Gamma$ -- A path. Both analyses were obtained for the FM001 ordering in the NdPt$_2$B crystal.}
        \label{F4A}
\end{figure}

Finally, in Figs. \ref{F3A} and \ref{F4A} are presented the zoomed $\Gamma$ -- A Berry curvature and $s_z$ spin-polarized electronic structure for the LaPt$_2$B and NdPt$_2$B, respectivelly. Here, it can be correlated the $\Omega$ values with the bands decoherence close to the Fermi level. 
As such, only a non-zero $\Omega_z$ component is observed as a result, in the LaPt$_2$B case, of the spitted spin-up and spin-down bands crossing the Fermi level.
In the NdPt$_2$B case, we observed the same phenomenon, however, a spin-polarized flat bands lie near the Fermi level providing the major contribution to $\Omega_z$.


\bibliography{library}

\end{document}